\documentclass[aps,prd,preprint,groupedaddress,showpacs,showkeys]{revtex4}
\usepackage{epsfig}
\begin{document}
\def\be{\begin{equation}} 
\def\ee{\end{equation}} 
\def\bea{\begin{eqnarray}} 
\def\eea{\end{eqnarray}} 
\newcommand{\nablaslash}{\nabla \hspace{-0.65em}/}
\title{The chiral condensate of strongly coupled QCD in the 't Hooft limit}

\author{G.Grignani}

\affiliation{Dipartimento di Fisica and Sezione I.N.F.N. 
Universit\'a di Perugia,
Via A. Pascoli,06123 Perugia, Italy}

\author{D.Marmottini}

\affiliation{Dipartimento di Fisica and Sezione I.N.F.N. Universit\'a di
Perugia, Via A. Pascoli,06123 Perugia, Italy}
\author{P.Sodano}

\affiliation{Dipartimento di Fisica and Sezione I.N.F.N. 
Universit\'a di Perugia,
Via A. Pascoli,06123 Perugia, Italy}

\date{\today}
\begin{abstract}
Using the recently proposed generalization to an arbitrary number of 
colors of the strong coupling approach to lattice gauge 
theories\cite{Grignani:2003uv},
we compute the chiral condensate of massless QCD in the 't Hooft limit. 
\end{abstract}



\maketitle

\section{Introduction}

As it is well known, the  spontaneous
breakdown of the chiral symmetry in massless QCD is signaled by 
the appearance of a non-vanishing order parameter, the chiral condensate 
$\chi_L=<\overline{\Psi}\Psi>$.
This quantity is intrinsically non perturbative and its determination is
not at all straightforward.
Since numerical computations on the lattice yield -at a given fixed lattice
spacing- only a bare result, one needs renormalization in order to use
the result for phenomenological applications. 
The results of several recent  
studies~\cite{Giusti:1998wy,Giusti:2001pk,Hernandez:1999cu,Hernandez:2001yn}
evidence by now the dependence of the chiral condensate on both the scale
and the renormalization scheme but allow for the computation of
a scale independent Renormalization Group Invariant (RGI) chiral
condensate~\cite{Giusti:2001pk}.
Recently, an interesting analytical evaluation of the quark condensate
in one-flavor massless QCD from the value of the gluino condensate in SUSY 
Yang-Mills theory by means of orientifold large-$N$ expansion 
has appeared in the literature~\cite{Armoni:2003yv}.

In this paper we compute the bare chiral condensate 
in the Hamiltonian formulation of lattice QCD using staggered fermions
in the strong coupling limit of the theory for any value of $N_C$;
the lattice chiral condensate is evaluated using $1/g^2$ 
as the expansion parameter and, then, taking the 't Hooft limit (large $N_C$,
with $g^2N_C$ fixed).
The extrapolation to the continuum limit is carried out by means of Pad\'e
approximants.
The tools used for our analysis were developed in~\cite{Grignani:2003uv};
here we shall present only the result for the chiral condensate referring
the reader to~\cite{Grignani:2003uv} for the details of the method.\\
The result obtained for the chiral condensate $\chi_L$ is compared 
to the recent numerical determination of the RGI chiral
condensate~\cite{Giusti:2001pk}.
In particular,
we estimate the scale independent and dimensionless ratio between $\chi_L$ and 
the cube of the mass of the $\rho$ meson, which has
been obtained in a strong coupling large $N_C$ calculation 
in~\cite{Grignani:2003uv}, 
and compare it to the numerical value of the ratio between 
the scale independent RGI chiral condensate obtained 
in~\cite{Giusti:2001pk}
%~\cite{Giusti:1998wy,Giusti:2001pk,Hernandez:1999cu,Hernandez:2001yn} 
%using different renormalization schemes
and the experimental value of the cube of the mass of the $\rho$ meson;
the check allows for an estimate of the lattice light velocity $t$
on the lattice.
$t$ turns out to be equal to $1.025$, in excellent agreement with 
the expected result of 1. This result in turn 
confirms the accuracy of our previous evaluation of the mass spectrum 
of mesons~\cite{Grignani:2003uv}.

The strong coupling limit, where the hadrons are automatically confined,
is a natural starting point for the study of the long-distance, 
non-perturbative features of the chiral symmetry breaking in massless QCD.
Even if, at weak coupling, the Wilson formalism yields an accurate analysis
of the $U(1)$ anomaly and PCAC~\cite{Karsten:1980wd}, one cannot use it
to compute the chiral condensate at strong coupling since it explicitly
breaks the chiral symmetry on the lattice.

The staggered fermion formalism yields, instead, a remnant of 
chiral symmetry on the lattice, which is the invariance of the theory under
translation by a single link; thus, a spontaneous breakdown of chiral 
symmetry may be evidenced also at strong coupling via a non vanishing
chiral condensate. 
Moreover, the staggered fermion formalism is known to yield good results 
in the strong coupling evaluation of the hadron
spectrum~\cite{Grignani:2003uv,Banks:1976ia,Grignani:2003ix} and
of chiral condensate of lower dimensional 
models~\cite{Berruto:1997jv,Berruto:1999cy}
Other types of lattice fermions such as domain-wall 
or overlap fermions are expected to suffer both 
doubling and explicit breaking of chiral symmetry\cite{Brower:1999ak}
at strong coupling.

Strongly coupled lattice gauge theories are intimately 
related to quantum spin systems~\cite{Smit:1980nf}; in particular lattice 
gauge theories with staggered fermions exhibit
interesting similarities with condensed matter systems.
For example, it is well known that the quantum spin-$1/2$ Heisenberg 
antiferromagnet is equivalent to the strong coupling limit of either a $U(1)$ 
or $SU(2)$ lattice gauge 
theory~\cite{Marston:gx,Berruto:1998tg,Berruto:1999ga}. 
For the gauge group $U({\mathcal N}_c)$, one may establish the equivalence 
with a spin-${\mathcal N}_C/2$ Heisenberg antiferromagnet~\cite{Semenoff:gz}. 
Since the seminal work of 't Hooft~\cite{'tHooft:1973jz,'tHooft:1974hx} 
the large $N_C$ limit (with $g^2 N_C$ 
fixed), has played an increasingly important role in studying gauge 
theories in the
continuum, on the lattice~\cite{Teper:2002kh,Grignani:2003uv} and 
the duality between gauge and string theories
~\cite{Aharony:1999ti,Kruczenski:2003be}.  
Furthermore, in a recent paper~\cite{Grignani:2003uv}, we have generalized 
the strong coupling calculation by Banks et al.~\cite{Banks:1976ia} to 
an arbitrary number of colors $N_C$ and we have  used the 't Hooft limit 
to investigate some features of the 
meson spectrum of strongly coupled lattice QCD.
Our results imply that, also at strong coupling, 
the 't Hooft limit offers a very accurate method to consistently
determine both the QCD spectrum and the chiral condensate.

\section{Strong coupling evaluation of the QCD chiral condensate in the 't Hooft limit}

In the Hamiltonian formulation of lattice QCD with staggered 
fermions~\cite{Susskind:1976jm} time is a continuous variable and space
is discretized on a 3-dimensional cubic lattice with $M$ sites, labeled
by $\vec r=(x,y,z)$; with $x$, $y$ and $z$ integers. 
The lattice Hamiltonian with one flavor of massless quark may be written 
as the sum of three contributions 
\be
\label{Ham}
H=H_e+\tilde H_q+H_m,
\ee
where
\bea
\label{H}
&& H_e=\frac{g^2}{2a}\sum_{[\vec{r},\hat{n}]}E^a[\vec{r},\hat{n}]^2\\
\label{I}
&&
\tilde{H}_q=\frac{t}{2a}\sum_{[\vec{r},\hat{n}]}\eta(\hat{n})
\Psi_A^{\dagger}(\vec{r}
+\hat{n})U_{AB}[\vec{r},\hat{n}]\Psi_B(\vec{r})+h.c.\equiv
H_q+H_q^{\dagger}\\
\label{L}
&&
H_m=\frac{1}{2g^2a}\sum_{[\vec{r},\hat{n},\hat{m}]}\left[Tr(U[\vec{r},
\hat{n}]U[\vec{r}+\hat{n},\hat m]U^{\dagger}[\vec{r}+\hat{m},\hat n]
U^{\dagger}[\vec{r},\hat{m}])+h.c.\right]
\eea
are the electric field Hamiltonian, the interaction Hamiltonian between 
quarks and gauge fields and the magnetic Hamiltonian, respectively.
$t$ is the lattice light velocity.
The sums $\sum_{[\vec{r},\hat{n}]}$ are extended to the $N$ lattice 
links, whereas $\sum_{[\vec{r},\hat{n},\hat{m}]}$ is a sum over the
plaquettes.
$\hat{n}=\hat{x},\hat{y},\hat{z}$ is the unit vector in the $\vec n$ direction 
and
\begin{equation}
\label{B}
\eta(\hat{x})=(-1)^z, \qquad \eta(\hat{y})=(-1)^x, \qquad
\eta(\hat{z})=(-1)^y
\end{equation}
\noindent 
are the Dirac $\vec\alpha$ matrices for staggered fermions
~\cite{Susskind:1976jm}.
The gauge field $U[\vec r,\hat n]$ is associated with the link 
$[\vec r,\hat n]$ ant it is a group element in the fundamental representation 
of $SU(N_C)$.
In the strong coupling expansion the electric field Hamiltonian $H_e$, 
(\ref{H}), 
is the unperturbed Hamiltonian while the interaction
Hamiltonian between quarks and gauge fields $\tilde H_q$, (\ref{I}), 
and the magnetic Hamiltonian $H_m$, (\ref{L}), are treated as perturbations.
An important feature of the Hamiltonian (\ref{Ham}) is its invariance under
translation by a single link which plays the role of a discrete chiral 
symmetry~\cite{Banks:1976ia}. 
It takes even sites into odd sites (even (odd) sites are those 
with $x+y+z$ even (odd)).
The pertinent lattice transformations have the form
$\Psi(r)\rightarrow\Psi(r+\hat{x})(-1)^y$, $\Psi(r)\rightarrow\Psi(r+\hat{y})(-1)^z$
and $\Psi(r)\rightarrow\Psi(r+\hat{z})(-1)^x$.
In momentum space the last equation can be written as
$q\to e^{ik_z}\gamma_{5}\tau_{3}q$,
which in the continuum limit, where $k_z$ is infinitesimal, becomes
$q\to \gamma_5\tau_3q$.
The other two transformations yield $q\to \gamma_5\tau_2q$, $q\to \gamma_5\tau_1q$.

$H_e$ has two degenerate
ground states corresponding to those of the spin $N_C/2$ antiferromagnetic
Ising model. One state may be obtained from the other
by interchanging odd and even sites;
choosing one of these two vacua leads to the spontaneous breakdown
of the chiral symmetry.
In the thermodynamic limit
the two ground states are not mixed to any finite order of perturbation
theory\cite{Berruto:1997jv} and thus, in the perturbative expansion, 
one only has to consider
diagonal matrix elements and, consequently, perturbation theory for
non-degenerate states.
The ground state energy has been evaluated in the strong 
coupling regime in~\cite{Grignani:2003uv} up to the fourth order in
the strong coupling expansion and for a generic value of $N_C$.\\
The order parameter evidencing discrete chiral symmetry breaking
is the mass operator 
$\mathcal M=\bar\psi(\vec r)\psi(\vec r)$
since it acquires a nonzero expectation value giving rise to the chiral 
condensate.
In the staggered fermion formalism the pertinent lattice operator is
given by
\be
\label{QCDmassop}
\mathcal{M}=-\frac{1}{Ma^3}\sum_{\vec r}(-1)^{x+y+z}\psi^\dagger_A(\vec r)
\psi_A(\vec r).
\ee
One has to evaluate the expectation value of $\mathcal M$
on the perturbed states $|p_0>$ generated by applying $\tilde H_q$ to the
ground state $|0>$. One has
$|p_0>=|0>+|p_0^{(1)}>+|p_0^{(2)}>$,
where
\bea
&& |p_0^{(1)}>=\frac{\Pi_0}{E_0^{(0)}-H_e}\tilde H_q|0>\\
&& |p_0^{(2)}>=\frac{\Pi_0}{E_0^{(0)}-H_e}\tilde H_q
\frac{\Pi_0}{E_0^{(0)}-H_e}\tilde H_q|0>+
\frac{\Pi_0}{E_0^{(0)}-H_e}H_m|0>.
\eea
The lattice chiral condensate is then given by
\be
\chi_L=
\frac{<0|\mathcal M|0>+<p_0^{(1)}|\mathcal M|p_0^{(1)}>
+<p_0^{(2)}|\mathcal M|p_0^{(2)}>}{<0|0>
+<p_0^{(1)}|p_0^{(1)}>+<p_0^{(2)}|p_0^{(2)}>}
\label{chil}
\ee
where
$$<,>=\left\{\prod_{[\vec{r},\hat{n}]}\int dU[\vec{r},\hat{n}]\right\}(,)$$
\noindent is the inner product in the full Hilbert space of the model.
$dU$ is the Haar measure on the gauge group manifold and $(,)$
the fermion Fock space inner product;
$\Pi_0$ is the projection operator projecting onto states
orthogonal to $|0>$.

$\chi_L$ may be more conveniently computed 
by constructing an eigenstate of $H_e$ and using it to evaluate the
functions $f(H_e)$ appearing in Eq.(\ref{chil}).
In order to show the method used to evaluate $\chi_L$~\cite{Grignani:2003uv} we will concentrate
on the denominator of Eq.(\ref{chil}).
Since the vacuum state
$|0>$ is a singlet of the electric field algebra, one has $E^a[\vec{r},\hat{n}]|0>=0$
which, in turn, implies that
$H_e|0>=0$. Using the left action of the Lie algebra
generated by the electric field $E^a[\vec r,\hat n]$ on $U[\vec r,\hat n]$,
$H_e|0>=0$ and putting the
commutator $[H_e,U[\vec{r},\hat{n}]]|0>$ in place of
$H_eU[\vec{r},\hat{n}]|0>$, one finds
\begin{equation}
\label{N}
H_eU[\vec{r},\hat{n}]|0>=\frac{g^2}{2a}C_2(N_C)U[\vec{r},\hat{n}]|0>,
\end{equation}
where $C_2=(N_C^2-1)/2N_C$ is the Casimir operator of
$SU(N_C)$. $U[\vec{r},\hat{n}]|0>$ is then an eigenstate of $H_e$ with
eigenvalue $g^2C_2(N_C)/2a$.
Consequently,
\be
\label{UU}
<0|\tilde{H}_q\frac{1}{(E_0-H_e)^2}\tilde{H}_q|0>=
\frac{8a^2}{g^4C^2_2}<0|{H}^\dagger_q{H}_q|0>.
\ee
Taking into account eq.(\ref{UU}) after integration over the link variable
$U$ \cite{Creutz:ub} and the fact that $<0|0>=1$, one finds 
\be
\label{gs}
<p_0^{(1)}|p_0^{(1)}>=\frac{t^2}{g^4C_2^2}N_CN.
\ee
To compute $<p_0^{(2)}|p_0^{(2)}>$ one needs to construct
suitable eigenstates of the unperturbed Hamiltonian $H_e$ containing 
two and four link variables
$U[\vec{r},\hat{n}]$~\cite{Grignani:2003uv}. Using the eigenvalues of the 
$f(H_e)$'s on these eigenstates one finds
\bea
\nonumber
<p_0^{(2)}|p_0^{(2)}>&=&
\frac{4a^2}{g^4C_2^2}\left[\frac{4a^2}{g^4C_2^2}<0|H_qH_q^{\dagger}
H_qH_q^{\dagger}|0>+
\frac{2a^2}{g^4C_2^2}\left(<0|H_qH_qH_q^{\dagger}
H_q^{\dagger}|0>\right.\right.\\
\nonumber
&+&\left.\left.\frac{2N_C-3}{(N_C-2)^2}<0|H_qH_q
\sum_{[\vec r,\hat n]}\Psi_A^\dagger(\vec r)U_{AB}^\dagger[\vec r,\hat n]
\Psi_B(\vec r+\hat n)\right.\right.\\
\nonumber
&\times&\left.\left.\Psi_C^\dagger(\vec r)U_{CD}^\dagger[\vec r,\hat n]
\Psi_D(\vec r+\hat n)|0>\right)\right]\\
\nonumber
&+&\frac{1}{4g^8C_2^2}<0|\sum_{[\vec r,\hat n,\hat m]}U_{AB}[\vec r,\hat n]
U_{BC}[\vec r+\hat n,\hat m]U_{CD}^{\dagger}[\vec r+\hat m,\hat n]
U_{DA}^{\dagger}[\vec r,\hat m]\\
&\times&\sum_{[\vec{r^{\prime}},\hat l,\hat k]}U_{EF}[\vec{r^{\prime}},
\hat l]U_{FG}[\vec{r^{\prime}}+\hat l,\hat k]U_{GH}^{\dagger}[\vec{r^{\prime}}
+\hat k,\hat l]U_{HE}^{\dagger}[\vec{r^{\prime}},\hat k]|0>
\eea
and the integration over the link variables leads to
\bea
&& <p_0^{(2)}|p_0^{(2)}>=\frac{t^4}{g^8C_2^4}\left[\left(-5N_C
-\frac{N_C^2}{2}\right)N+
\frac{N_C^2}{2}N^2+\frac{N_C(N_C-1)^3}{2(N_C-2)^2}\right]\\
\label{gs1}
&& \phantom{<p_0^{(2)}|p_0^{(2)}>=}+\frac{N}{g^8C_2^2}.
\eea
The expectation values of the mass operator $\mathcal M$ are computed
using a similar procedure and they are given by
\bea
&& <0|\mathcal M|0>=-\frac{N_C}{2a^3}\\
&& <p_0^{(1)}|\mathcal M|p_0^{(1)}>=-\frac{t^2}{g^4C_2^2a^3}
\left[-6N_C+\frac{N_C^2}{2}N\right]\\
\nonumber
&& <p_0^{(2)}|\mathcal M|p_0^{(2)}>=-\frac{t^4}{g^8C_2^4a^3}\left[60N_C+6N_C^2
-\frac{17}{2}N_C^2N-\frac{N_C^3}{4}N\right.\\
\label{mexp}
&& \left.+\frac{N_C^3}{4}N^2+\frac{N_C^3}{4}N^2+\frac{N_C(N_C-1)^3}{(N_C-2)^2}
(-6+\frac{N_C}{4}N)\right]-\frac{N_C}{2a^3}\frac{N}{g^8C_2^2} .
\eea
The last contributions in eq.(\ref{gs1}) and in eq.(\ref{mexp}) 
come from the magnetic term in $|p_0^{(2)}>$.
Using eqs.(\ref{gs}-\ref{mexp}),
one finds a non-vanishing chiral condensate for any finite $N_C$.
For brevity, we do not write the expression
of $\chi_L$ for a generic value of $N_C$, since it may be easily inferred
from (\ref{chil}-\ref{mexp}).
It is easy to see that, as it should be, 
all the dependence on the number of lattice links $N$
disappears from the evaluation of $\chi_L$  
up to the order $t^4/g^8$  and this is a very non trivial 
check of our calculations.\\
Consider now the 't Hooft limit
where $g^2N_C$ is rescaled to $g^2$ and $N_C$ is then sent to 
infinity. One gets for the bare lattice chiral condensate
\be
\label{QCDchircond}
\chi_L=-\frac{N_C}{a^3}\left[\frac{1}{2}-24\epsilon+864\epsilon^2\right],
\ee
where $\epsilon=t^2/g^4$.

\section{Comparison with the RGI chiral condensate}

To compare the results of the strong coupling expansion
with the numerical value of the scale independent RGI chiral condensate determined 
in~\cite{Giusti:2001pk} one needs to construct a scale independent
quantity. This can be done by considering for example the ratio between
$\chi_L$ which has dimension $($mass$)^3$ and the mass cube of a meson.
In a previous work~\cite{Grignani:2003uv} we derived, in strong coupling, the series 
expansions for the masses of the low-lying states in the meson spectrum
obtaining a good agreement between the experimental values for the 
meson mass ratios and our lattice results.
For the $\rho$ meson we found~\cite{Grignani:2003uv}
\be
\label{rho}
m_{\rho}=\frac{g^2}{a}\left[\frac{1}{4}+6\epsilon-203\epsilon^2\right].
\ee
In order to eliminate the 
dependence on $a^3$, one may then consider the ratio between $\chi\equiv -\chi_L/N_C$ 
and $m_{\rho}^3$
\be
\frac{\chi}{m_{\rho}^3}=\frac{1}{g^6}
\frac{\frac{1}{2}-24\epsilon+864\epsilon^2}
{\left(\frac{1}{4}+6\epsilon-203\epsilon^2\right)^3}.
\label{ratio}
\ee
To compare this strong coupling result with the numerical value of the RGI
chiral condensate, the series derived in the strong
coupling regime ($\epsilon=t^2/g^4 \ll 1$), need to be extrapolated
to the region in which $\epsilon \gg 1$, which corresponds to weak coupling.
To make this extrapolation possible it is customary to use the Pad\'e 
approximant method which allows one to extrapolate a series beyond its
convergence radius.
Using
$1/g^6=\epsilon^{3/2}/t^3$,
from (\ref{ratio}) one gets
\be
\left[\frac{\chi}{m_{\rho}^3}\right]^{4/3}=\frac{\epsilon^2}{t^4}
\left[\frac{\frac{1}{2}-24\epsilon+864\epsilon^2}
{\left(\frac{1}{4}+6\epsilon-203\epsilon^2\right)^3}\right]^{4/3}.
\ee
Using the $[0,2]$ Pad\'e approximant, one has
\be
\label{chi/rho}
\left[\frac{\chi}{m_{\rho}^3}\right]^{4/3}\simeq
\frac{\epsilon^2}{t^4}\frac{2^{2/3}\times 64}{1+160\epsilon+7632\epsilon^2}.
\ee
For ($\epsilon\to\infty$), eq.(\ref{chi/rho}) becomes
\be
\label{chi/rho cont}
\left[\frac{\chi}{m_{\rho}^3}\right]^{4/3}\stackrel{\epsilon\to\infty}
{\rightarrow}\left(\frac{2^{2/3}\times 64}{7632}\right)\frac{1}{t^4}.
\ee
If one takes for the chiral condensate $\chi$ the RGI numerical value obtained
in~\cite{Giusti:2001pk}
%~\cite{Giusti:1998wy,Giusti:2001pk,Hernandez:1999cu,Hernandez:2001yn} 
and for the value of the mass of the $\rho$ meson its experimental value,
equating their ratio to (\ref{chi/rho cont}), one obtains
\be
\label{chi/rho lat}
\left(\frac{2^{2/3}\times 64}{7632}\right)\frac{1}{t^4}=
\left[\frac{(\chi_{\rm {num}}\pm \Delta \chi_{\rm {num}})}
{((0.771\pm 0.0009))^3}
\right]^{4/3},
\ee
where $\chi_{\rm {num}}$ is the numerical value of the chiral condensate
of ref.~\cite{Giusti:2001pk}
\be
\chi_{\rm {num}}\pm \Delta\chi=(0.0167 \pm 0.0031)\rm {Gev}^3
\ee 
>From eq.(\ref{chi/rho lat}) one gets for $t$ the following value
\be
t=1.025 \pm 0.177
\ee

The result obtained is close
to the expected value of 1, within the error, thus showing very good
agreement between our results for the ratio $\chi_L/m_\rho^3$ and the one
obtained from the results of~\cite{Giusti:2001pk}.
%~\cite{Giusti:1998wy,Giusti:2001pk,Hernandez:1999cu,Hernandez:2001yn}.
In the r.h.s. of eq.(\ref{chi/rho lat}) the error on the numerical value
of $\chi_L$ is due to statistical effects
while the error on the mass of $\rho$ is experimental.
Of course, for lattice QCD we can only check the consistency of the strong
coupling determination of the chiral condensate by comparison of our 
result with the numerical value of the RGI chiral condensate determined 
in~\cite{Giusti:2001pk}.
A comparison with the ``true value''of the continuum chiral condensate
is -for QCD- not only impossible but also senseless due to the manifest scale 
and renormalization scheme dependence of the lattice determinations of this 
quantity~\cite{Giusti:1998wy,Giusti:2001pk,Hernandez:1999cu,Hernandez:2001yn}.
As we shall see immediately after, in the simpler case of the Schwinger
model, the strong coupling evaluation of the scale independent ratio
between the chiral condensate and the pseudoscalar excitation mass matches
very well the exact continuum value.

\section{Concluding remarks}

The reliability of the method used to estimate the 
accuracy of our computation of the chiral 
condensate in QCD may be usefully tested in the one flavor Schwinger
model for which the value of the chiral condensate and the mass of the
pseudoscalar boson in the continuum are known exactly.
In fact, one may compute the ratio between the lattice chiral condensate 
and the lattice pseudoscalar mass and then equate it to its continuum exact
counterpart.
For the sake of clarity, we briefly recall the results of Berruto 
et al.~\cite{Berruto:1997jv} for the evaluation of the lattice chiral 
condensate and the pseudoscalar mass of the strongly coupled one-flavor 
Schwinger model in the Hamiltonian approach with staggered fermions. 
In the continuum gauge model the chiral symmetry is broken by the anomaly.
In the Hamiltonian lattice formulation of the Schwinger model the 
axial symmetry is spontaneously broken 
via a non zero expectation value of the chiral condensate.
In the continuum theory it is well known that~\cite{Nielsen:1976hs}
$<\bar \psi(x)\psi(x)>=-e^{\gamma}e_c/(2\pi\sqrt\pi)$,
where $\gamma=0.577...$ is the Euler constant and $m=e_c/\sqrt{\pi}$ is the 
mass of the pseudoscalar excitation.\\
A lattice Hamiltonian which,
in the continuum limit, reduces to the Schwinger Hamiltonian
is
\be
\label{HSchw}
H_S=\frac{e_L^2a}{2}\sum_{x}E_x^2-\frac{it}{2a}\sum_{x}(\psi^\dagger_{x+1}
e^{iA_x}\psi_x-\psi^\dagger_{x}e^{-iA_x}\psi_{x+1})\equiv H_u+H_p,
\ee
where the fermion fields are defined on the sites, $x=-\frac{N}{2},
-\frac{N}{2}+1,...,\frac{N}{2}$, gauge and electric fields, $A_x$ and $E_x$,
on the links $[x,x+1]$; $N$ is an even integer.
The coefficient $t$ of the hopping term in (\ref{HSchw}) plays the role of 
the lattice light speed. In the naive continuum limit, $e_L=e_c$ and $t=1$.\\
In the strong coupling limit the electric field Hamiltonian $H_u$ is the 
unperturbed Hamiltonian while the hopping Hamiltonian $H_p$ is treated as
a perturbation.
As in 3+1 dimensional QCD, there are two degenerate gauge invariant 
ground states which, in the Coulomb gauge, have the form
\be
|\psi>=\prod_{x=even}\psi^\dagger_x|0>, \qquad \qquad \qquad 
|\chi>=\prod_{x=odd}\psi^\dagger_x|0>.
\ee
To the fourth order in $\alpha=t/2e_L^2a^2$, the perturbative expansion 
for the ground state energy is given by~\cite{Berruto:1997jv}
\be
\label{ES}
E_\psi=E_\psi^{(0)}+\alpha^2E_\psi^{(2)}+\alpha^4E_\psi^{(4)}=
\frac{N}{32}-4N\alpha^2+192N\alpha^4.
\ee
On the lattice the pseudoscalar boson of the continuum Schwinger model
is provided by the operator
$|\theta>=1/(\sqrt N)\sum_{x=1}^{N}(\psi_x^\dagger e^{iA}\psi_{x+1}
+\psi_{x+1}^\dagger e^{-iA}\psi_{x})$.
The energy of this state was computed up to the fourth order in the 
strong coupling expansion in~\cite{Berruto:1997jv} and its mass is 
given by subtracting the ground state energy (\ref{ES}) from it
\be
\label{pseudmass}
m_p=E_\theta-E_\psi=e_L^2a(\frac{1}{4}+8\alpha^2-576\alpha^4).
\ee
The lattice chiral condensate in the staggered fermion formalism may be
obtained by considering the mass operator
$M(x)=-1/(Na)\sum_{x=1}^N(-1)^x\psi_x^\dagger\psi_x$
and evaluating its expectation value on the perturbed state $|p_\psi>$
generated by applying $H_h$ to $|\psi>$, 
$|p_\psi>=|\psi>+|p_\psi^{(1)}>+|p_\psi^{(2)}>$.
A direct computation of the lattice chiral
condensate is given in~\cite{Berruto:1997jv} and yields
\be
\label{Schircond}
\chi_L =\frac{<p_\psi|M|p_\psi>}{<p_\psi|p_\psi>}=
-\frac{1}{a}\left(\frac{1}{2}-32\alpha^2+1536\alpha^4\right).
\ee
We here compute the ratio between the lattice chiral condensate 
(\ref{Schircond}) and the lattice pseudoscalar mass (\ref{pseudmass}) 
and equate it to its continuum value
\be
\label{Scond/mass}
-\left(\frac{2}{e_L^2a^2}\right)
\frac{\frac{1}{4}-16z+768z^2}{\frac{1}{4}+8z-576z^2}=-\frac{e^\gamma}{2\pi},
\ee
where $z=\alpha^2=\frac{t^2}{4e_L^4a^4}$. Eq.(\ref{Scond/mass}) 
is true only when 
Pad\'e approximants are used, since the l.h.s. holds only for $z\ll 1$,
while the r.h.s. provides the value of the ratio between the chiral 
condensate and the pseudoscalar mass obtained when $z\cong\infty$.
Using the fourth power of eq.(\ref{Scond/mass}) in order to
construct the $[0,2]$ Pad\'e approximant for the l.h.s. 
of eq.(\ref{Scond/mass}), one gets
\be
\left(\frac{\chi_L}{m_p}\right)^4=\frac{256}{t^4}\frac{z^2}
{1+384z+58368z^2}.
\ee
One may now take the limit $z\to\infty$ obtaining
$\left(\chi_L/m_p\right)^4=1/(228t^4)$ and, 
equating it to its continuum counterpart, one gets an
equation for $t$
\be
\label{St}
\frac{1}{228t^4}=\left(\frac{e^\gamma}{2\pi}\right)^4.
\ee
>From eq.(\ref{St}) this one gets $t=1.03$ for the lattice light velocity;
the result lies $3\%$ above the exact value thus showing that the 
strong coupling evaluation of the chiral condensate yields a very good 
result also for this model.

In conclusion,
our large $N_C$ Hamiltonian approach with staggered fermions evidences
that the possible ground state of strongly coupled lattice QCD are those
of a spin $N_C/2$ antiferromagnetic Ising model; choosing one of the
two ground states amounts then to the spontaneous breaking of the discrete
chiral symmetry corresponding to translations by a lattice site.
As a consequence a non vanishing chiral condensate is found which is 
the order parameter for the spontaneous breakdown of the chiral symmetry.
The bare lattice chiral condensate $<\bar\psi\psi>$ is computed in the 
strong coupling region for a general number of colors $N_C$ 
using the method described in~\cite{Grignani:2003uv}. The result obtained is 
intensive, i.e. independent on the number of lattice links.
Then the 't Hooft limit is taken;
the ratio between the chiral condensate, obtained in this way, 
and the third power of the $\rho$ meson mass computed in the 
't Hooft limit in~\cite{Grignani:2003uv} is used to provide an
estimate of the lattice light velocity, $t$. $t$ turns out to
be very close to the expected value of 1;
this evidences that the large $N_C$ limit is a very reliable and relatively 
simple pathway for the evaluation of the chiral condensate also in 
the strong coupling region.

% \appendix

% \section{}
% \label{}

\end{document}